\documentclass{aa}  
\usepackage{graphicx}
\usepackage{aas_macros}
\usepackage{ulem}
\usepackage{epstopdf}
\usepackage{amsmath}
\usepackage{amssymb}
\usepackage{mathtools}
\usepackage{mathrsfs}
\usepackage[switch]
{lineno}
\usepackage{bm}
\usepackage{color}
\usepackage{xcolor}
\usepackage{textcomp}
\usepackage{gensymb}
\usepackage{hyperref}
\usepackage{txfonts}
\newcommand{\isois}{IS\(\odot\)IS}
\newcommand{\delbx}{\delta b_x}
\newcommand{\delby}{\delta b_y}
\newcommand{\delbz}{\delta b_z}
\begin{document}

   \title{Magnetic Field Line Random Walk and Solar Energetic Particle Path Lengths}

   \subtitle{Stochastic Theory 
   and PSP/IS\(\odot\)IS Observations}

   \author{R. Chhiber
          \inst{1,2}
          \and
          W. H. Matthaeus\inst{1}
          \and
          C.M.S. Cohen\inst{3}
          \and
          D. Ruffolo\inst{4}
          \and
          W. Sonsrettee\inst{5}
          \and
          P. Tooprakai\inst{6}
          \and
          A. Seripienlert\inst{7}
          \and
          P. Chuychai\inst{8}
          \and
          A. V. Usmanov\inst{1,2}
                   \and
          M. L. Goldstein\inst{10}
          \and
          D. J. McComas\inst{9}
          \and
          R. A. Leske\inst{3}
          \and
          E. R. Christian\inst{2}
          \and
          R. A. Mewaldt\inst{3}
          \and
          A. W. Labrador\inst{3}
          \and
          J. R. Szalay\inst{9}
          \and
          C.~J. Joyce\inst{9}
          \and
          J. Giacalone\inst{11}
          \and
          N.~A. Schwadron\inst{12}
          \and
          D.~G. Mitchell\inst{13}
          \and
          M.~E. Hill\inst{13}
          \and
          M.~E. Wiedenbeck\inst{15}
          \and
          R.~L. McNutt Jr.\inst{13}
          \and
          M. I. Desai\inst{14}
     }

   \institute{Department of Physics and Astronomy and Bartol Research Institute, University of Delaware, Newark, DE 19716, USA\\
            \email{rohitc@udel.edu}
         \and
             Heliophysics Science Division, NASA Goddard Space Flight Center, Greenbelt MD 20771, USA\\
             \email{rohit.chhiber@nasa.gov}
    \and
    California Institute of Technology, Pasadena, CA 91125, USA
    \and
    Department of Physics, Faculty of Science, Mahidol University, Bangkok 10400, Thailand
    \and
    Faculty of Engineering and Technology, Panyapiwat Institute of Management, Nonthaburi 11120, Thailand
    \and
    Department of Physics, Faculty of Science, Chulalongkorn University, Bangkok 10330, Thailand
    \and
    National Astronomical Research Institute of Thailand (NARIT), Chiang Mai 50180, Thailand 
    \and
    33/5 Moo 16, Tambon Bandu, Muang District, Chiang Rai 57100, Thailand
    \and
    Department of Astrophysical Sciences, Princeton University, Princeton, NJ 08544, USA
    \and
    University of Maryland Baltimore County, Baltimore, MD 21250, USA
    \and
    University of Arizona, Tucson, AZ 85721, USA
    \and
    University of New Hampshire, Durham, NH, 03824, USA
    \and
    Johns Hopkins University Applied Physics Laboratory, Laurel, MD 20723, USA    
    \and
    University of Texas at San Antonio, San Antonio, TX 78249, USA
    \and
    Jet Propulsion Laboratory, California Institute of Technology, Pasadena, CA 91109, USA
}

   \date{Received October --, 2020; accepted -- --, 2020}

 
  \abstract
{In 2020 May-June, six solar energetic ion events were observed by the Parker Solar Probe/\isois~instrument suite at $\approx$0.35 AU from the Sun.  
From standard velocity-dispersion analysis, the apparent ion path length is $\approx$0.625 AU at the onset of each event.}
{We develop a formalism for estimating the path length of random-walking magnetic field lines, to explain why the apparent ion path length at event onset greatly exceeds the radial distance from the Sun for these events.}
{We developed analytical estimates of the average increase in path length of random-walking magnetic field lines, relative to the unperturbed mean field. 
Monte Carlo simulations of field line and particle trajectories in a model of solar wind turbulence are used to validate the formalism and study the path lengths of particle guiding-center and full-orbital trajectories. The formalism is implemented in a global solar wind model, and results are compared with ion path lengths inferred from \isois~observations.}
{Both a simple estimate and a rigorous theoretical formulation are obtained for field-lines' path length increase as a function of path length along the large-scale field. 
From simulated field line and particle trajectories, we find that particle guiding centers can have path lengths somewhat shorter than the average field line path length, while particle orbits can have substantially larger path lengths due to their gyromotion with a nonzero effective pitch angle.} 
  {The long apparent pathlength during these solar energetic ion events can be explained by 1) a magnetic field line path length increase due to the field line random walk, and 2) particle transport about the guiding center with a nonzero effective pitch angle.  
   Our formalism for computing the magnetic field line path length, accounting for turbulent fluctuations, may be useful for application to solar particle transport in general.}


\maketitle
%
\section{Introduction}

The propagation of energetic particles in the solar wind or other space and astrophysical plasmas is a complex problem that involves scattering theory, as well as a quantitative understanding of both the  large-scale magnetic field and its turbulent fluctuations \citep{fisk1978interactions,shalchi2009}. Taken together, these magnetic field properties are responsible for particle transport. Besides causing pitch-angle scattering and parallel diffusion, magnetic fluctuations also contribute in a fundamental way to perpendicular transport of particles, by deflecting the magnetic field lines in a random way, in a process often called {\it magnetic field line random walk.} or simply  ``FLRW'' \citep{jokipii1966cosmic,jokipii1969ApJstochastic}. Here we consider a specific effect of FLRW that is likely to be of particular importance for solar energetic particle (SEP) propagation, namely the increase in path length along the magnetic field due to random fluctuations. Path length is germane to the SEP problem because it is a factor in determining the arrival time of particles at a detector when the field line is mapped back to its apparent source in the lower solar atmosphere. While a precise evaluation of the field line length involves information specific to the case at hand, it turns out that there are general estimates that can be made based on simple assumptions about the magnetic field and the fluctuations that cause the FLRW. Following some background discussion in Section \ref{sec:background}, such a simple estimate is provided in Section \ref{sec:formulation}, based on an analytical treatment of the path length for a particular class of turbulent fluctuations that is approximately realized in the solar wind. 
In Section 4.1, we confirm that this theory can explain the average magnetic field line path length for a simulated two-component turbulent field used to model solar wind turbulence.
In Section 4.2, we also compare those results with the average path length of particle guiding-center and full-orbit trajectories.
In Section \ref{sec:results1}, we estimate the average magnetic field path length in the  context of a global heliospheric simulation with a turbulence transport model. Section \ref{sec:results2} applies these results to a set of SEP events observed by Parker Solar Probe (PSP) in its fifth orbit, using observations from the EPI-Hi instrument aboard the \isois~suite. We conclude with discussion in Section \ref{sec:conclude}. 

\section{Background and Context: Particles Following Field Lines}\label{sec:background}

Soon after the original formulation of 
the theory of magnetic field line random walk \citep{jokipii1966cosmic}, 
the idea was applied to understanding how charged particles escape from 
the galaxy by following magnetic field lines \citep{jokipii1969ApJstochastic}.
The fundamental assumption is that when magnetic field lines randomly 
meander out of the galaxy, so too will energetic 
particles because their gyrocenters on average follow the field lines. This is the so-called ``FLRW limit'' of particle transport. 
 
There are two complications to this simple picture. One is that the topology of the field lines and magnetic flux surfaces \citep{taylor1971PoF,kadomtsev1979ppcf,isichenko1991PPCF} might induce nonstandard transport regimes, including both superdiffusive and trapped field line behavior \citep{ruffolo2003ApJ,chuychai2007ApJ}. Adding particles to the field lines, these effects can give rise to both unexpectedly large transverse displacements, as well as local temporary trapping of particles that delays the approach to a fully diffusive limit \citep{tooprakai2007GRL,tooprakai2016ApJ}. 

Another complication is that parallel scattering of charged particles introduces a range of possible effects on perpendicular transport, including subdiffusion \citep[one type of which is known as ``compound diffusion'';][]{lingenfelter1971ApL,urch1977APSS}.
In simple terms, if a particle  is assumed to follow a well-defined  field line, then if resonant scattering causes a reversal of the particle direction, it will unravel the same perpendicular displacement that it accumulated in the earlier part of the trajectory. A major factor that controls whether or not this occurs \citep{qin2002ApJ,qin2002GRL} is whether the three-dimensional magnetic field admits sufficient spatial complexity in the cross-field direction. Again, in simple terms -- particles have finite gyroradius so they actually are following not one, but a bundle of field lines. If all the circumscribed field lines are parallel to one another, then the retracing of paths by particles establishes subdiffusion. But if the field lines differ sufficiently, the particles return along distinct field lines, and diffusion can be recovered \citep{qin2002GRL}. This the basis of Nonlinear Guiding Center  Theory (NLGC) and its variations \citep{matthaeus2003nonlinear,shalchi2010ApJ,ruffolo2012random}. The relationship between the FLRW particle transport regime, the compound subdiffusion regime, and the NLGC 
transport regime is an interesting one \citep{bieber1997perpendicular,kota2000ApJ,qin2002ApJ}, and the boundaries separating these regimes 
remain incompletely defined. For example, it is clear that the heuristic expectation that lower-energy particles necessarily follow field lines more precisely than higher energy particles sometimes breaks down due to the particles' contrasting parallel mean free paths, and the degree of transverse complexity of the turbulence  \citep{minnie2009JGR}. 

The upshot of this background in scattering physics is that it is not {\it a priori} obvious how to characterize the relationship between SEP transport and the lengths of trajectories of individual field lines. A complex set of issues enters, involving field-line topology, resonant power that may induce parallel scattering, transverse complexity of the turbulence (including its topology and critical points), and possibly other factors. The bottom line is that, based on present knowledge, it is not possible to state, for a given injection event with a range of energies, whether
and how closely particles will 
follow field lines from source to point of observation. Nevertheless it is entirely clear that the magnetic field and its own random trajectories will play some role, and almost certainly an important one, in controlling the paths taken by an ensemble of energetic particles in their transport from source to observation. Understanding the potential complexity of this question, it is difficult to assert that the random character of the magnetic field exerts negligible influence on the particle trajectories or the total path length that particles follow leading to their detection. On the other hand, the same multiplicity of factors involved makes it doubtful that we can a formulate a unique or precise answer to the question of SEP path lengths. Here we present a first effort at estimating the increased path lengths that particles experience due to the change in field line length induced by the classic field line random walk. Our perspective is that even if details of the particle motions are not known, the path length of the field lines sets a scale for determining the paths traversed by the particles. 

\section{Analytical Path-length Formulation for Random-walking Field Lines}\label{sec:formulation}

\begin{figure}
\centering
\includegraphics[scale=.2]{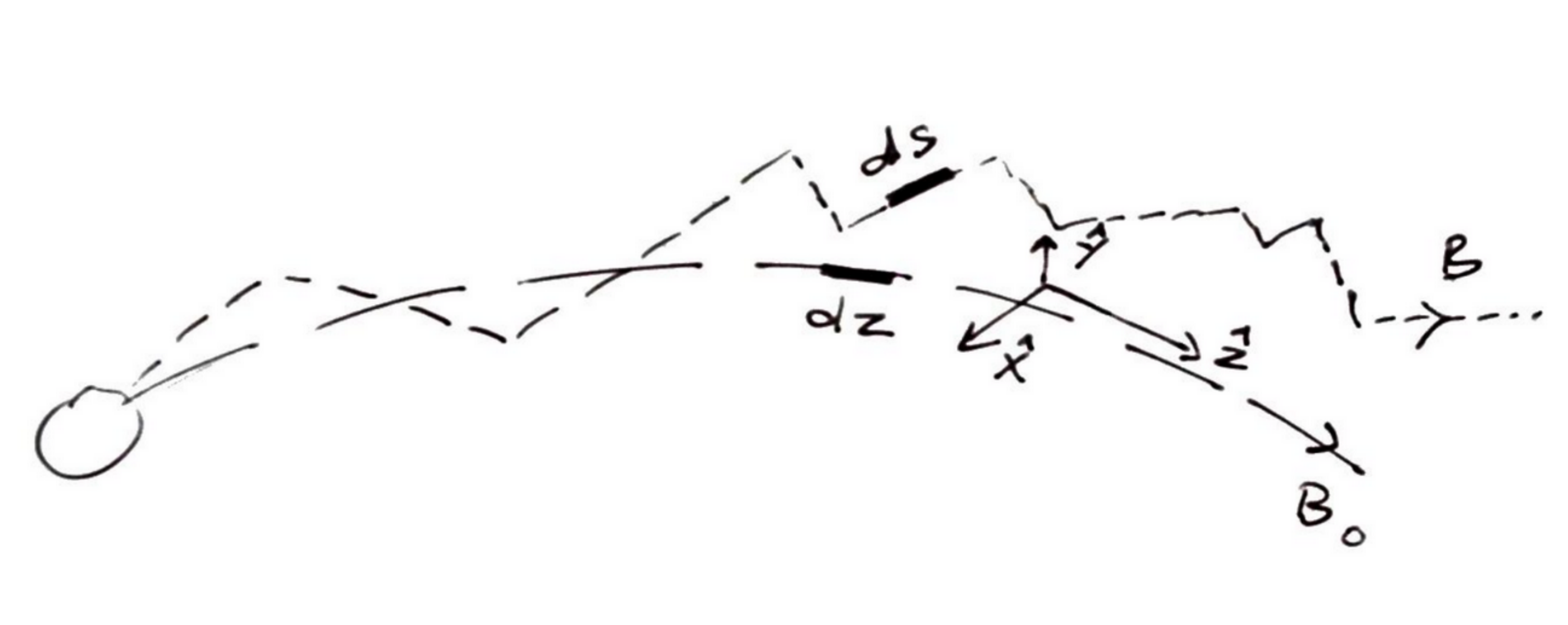}
\caption{Schematic showing central and random-walking field lines emerging from the Sun, and the local coordinate system used. Note that the magnetic field is statistically axisymmetric about the mean-field direction \(\hat{\bm{z}}\).}
\label{fig:cartoon}
\end{figure}

For a mean field \(B_0\hat{\bm{z}}\), a random-walking field line can be described by the equation
\begin{equation}
    \frac{ds}{B} = \frac{dz}{B_z} = \frac{dx}{b_x} = \frac{dy}{b_y},
    \label{eq:app_fl}
\end{equation}
where \(ds\) is the differential line element along the field line, \(B_z = B_0 + b_z\) is the \(z\)-component of the magnetic field, \(b_x,~b_y\), and \(b_z\) are the fluctuating components of the magnetic field in the \(x,~y,\) and \(z\) directions respectively, and \(B \equiv (b_x^2 + b_y^2 + B_z^2)^{1/2}\) is the magnitude of the magnetic field. We wish to estimate the path length of a random-walking field line in comparison with the slowly-varying central field line (which may be considered to be Parker-spiral-like). This path length  is given by integrating Equation \eqref{eq:app_fl} along the field line:
\begin{equation}
 S(r) =    \int ds = \int \frac{B}{B_z} dz.
    \label{eq:app_fl2}
\end{equation}
Note that the \(x,~y\), and \(z\) directions comprise a locally-defined 
coordinate system in which the \(z\)-direction is aligned with the central field-line, which has magnetic field strength \(B_0\) (see Figure \ref{fig:cartoon}).

%


It will be convenient to assume an implicit scale-separation between the large-scale field \(B_0\) and the fluctuating field, the latter varying in space much more rapidly than \(B_0\). To investigate the average behavior of the path length, we take an ensemble average of Equation \eqref{eq:app_fl2} over the ensemble of random-walking field lines, while integrating over the slowly varying (and non-random) spatial dependence of magnetic field properties:
\begin{equation}
     \langle S\rangle = \int \left\langle \frac{ B}{B_z} \right\rangle dz.
    \label{eq:app_fl3}
\end{equation}

\subsection{Simple Estimates of Path Length}\label{sec:formulation1}
To get an initial estimate of the path length, we replace the 
random magnetic fluctuations by 
a coherent estimate using their variances, enabling us
to write%
\begin{align}
    \left\langle \frac{ B}{B_z} \right\rangle &\sim \frac{[(B_0 \pm \delta b_z)^2 + \delta b_x^2 + \delta b_y^2]^{1/2}} 
    {B_0 \pm \delbz} \label{eq:app_ratio0}
    \\ &= 
    \left[1 + 
    \frac{\delta b_x^2 + \delta b_y^2}
    {(B_0 \pm \delbz)^2} \right]^{1/2},    \label{eq:app_ratio1}
\end{align}
where \(\delbx^2,~\delby^2\), and \(\delbz^2\) are the 
ensemble 
variances (or mean-squared fluctuations) of 
each of the cartesian 
components in 
\(x,~y\), and \(z\) directions. The \(\pm\) sign in Equations \eqref{eq:app_ratio0} and \eqref{eq:app_ratio1} accounts crudely for positive/negative \(\delbz\). 
 We define \(\delta b \equiv (\delbx^2 + \delby^2 + \delbz^2)^{1/2}\), and,  following observations \citep{bruno2013LRSP} and modeling \citep{chhiber2019psp2} of the inner heliosphere, in the remainder of the present section we assume that the turbulence is strong:
\begin{equation}
  \delta b/B_0 \sim 1.
  \label{eq:app_delbB}  
\end{equation}
We now evaluate Equation \eqref{eq:app_ratio1} for three different cases.

First we consider fully isotropic turbulence, which is a fundamental model that may be relevant, at least as a first approximation, when $\delta b \gg B_0$ such as in certain plasma regions within the magnetosheath or heliospheric current sheet. In that case polarity reversals will be difficult to avoid and different approaches may be advantageous \citep[see, e.g., ][]{Sonsrettee2015ApJ,sonsrettee2016ApJS}. Nevertheless we include a simple estimate for this case for context. Such a model may also be applicable downstream of high Mach number shocks, or in astrophysical settings such as the galactic halo \citep[][and references therein]{subedi2017ApJ}. 
For the case of isotropic variances, we have \(\delbx = \delby = \delbz\), and so Equation \eqref{eq:app_delbB} gives \(\delbz/B_0 \sim 1/\sqrt{3}\). Therefore,
\begin{align}
    \left\langle \frac{ B}{B_z} \right\rangle \sim 
    \left[1 + 
    \frac{2\delta b_z^2}
    {(B_0 \pm \delbz)^2} \right]^{1/2} 
    &= \left[1 + 
    \frac{2 (\delbz/B_0)^2}
    {(1 \pm \delbz/B_0)^2} \right]^{1/2}  \label{eq:ratio_iso1}
    \\
    &= 1.13, 2.18.  \label{eq:ratio_iso2}
\end{align}
The two estimates in Equation \eqref{eq:ratio_iso2} correspond to the positive and negative \(\delbz\) cases, respectively.

For variances in the ratio 5:4:1, a rough but often-quoted approximation relevant to typical solar wind observations \citep{belcher1971JGR}, we have \(\delbx^2 = 5\delbz^2\) and \(\delby^2 = 4\delbz^2\). Equation \eqref{eq:app_delbB} gives \(\delbz/B_0\sim 1/\sqrt{10}\). Then it follows from Equation \eqref{eq:app_ratio1} that
\begin{equation}
    \left\langle \frac{B}{B_z} \right\rangle \sim
    \left[1 + 
    \frac{9 (\delbz/B_0)^2}
    {(1 \pm \delbz/B_0)^2} \right]^{1/2} = 1.23, 1.71.
    \label{eq:ratio_541}
\end{equation}

For purely transverse fluctuations, i.e., Alfv\'en mode, a model of relevance in the corona and in reduced MHD contexts \citep{montgomery1982PhysScrip,rappazzo2008ApJ,oughton2017ApJ}, one specializes to the case \(\delbx = \delby\) and \(\delbz=0\). Equation \eqref{eq:app_delbB} reduces to \(\delbx/B_0\sim 1/\sqrt{2}\), and Equation \eqref{eq:app_ratio1} gives
\begin{equation}
    \left\langle \frac{B}{B_z} \right\rangle \sim
    \left[1 + \frac{2\delta b_x^2}{B_0^2} \right]^{1/2} 
    = 1.41 \label{eq:ratio_trans}
\end{equation}
%


At this stage, using the constant estimates of \(\langle B/B_z\rangle\) derived above with Equation \eqref{eq:app_fl3}, the ensemble-average path length for random-walking field lines can be estimated as \(\langle S\rangle = \langle B/B_z\rangle S_0\), where \(S_0 \equiv \int dz\) is the path length of the central, unperturbed field line (see Figure \ref{fig:cartoon}). We therefore see that, for \(\delta b/B_0=1\), the ratio \(S/S_0\) can be estimated as 1.13-2.18, 1.23-1.71, and 1.41, for the cases of isotropic, 5:4:1, and transverse fluctuations, respectively. Since magnetic fluctuations in the inner heliosphere are mainly transverse, these estimates suggest that, for a central field-line path length of about 1 AU, the average path length of a random walking field line is about \(\sim 1.2 - 1.7\) AU. The path length for field lines random walking in strong turbulence is then only fractionally larger from the Parker spiral length, according to these crude estimates. Note that our lower estimate of 1.2 is similar to the recent observational estimates in \cite{zhao2019ApJ}.

\subsection{Rigorous Estimate of Path Length}\label{sec:formulation2}
For a more rigorous estimate, we consider the case of uncorrelated and Gaussian-distributed fluctuations that are purely transverse to the mean-field direction. The probability distribution of the magnetic field magnitude \(B\) is then given by \citep[see][]{hartlep2000JGR}:
\begin{equation}
    f(B; B_0; \sigma_\perp^2) = \begin{cases}
     0 & B < B_0, \\
     \frac{B}{\sigma_\perp^2} \exp{\left(\frac{B_0^2-B^2}{2\sigma_\perp^2}\right)} 
     & B \geq B_0,
    \end{cases}
\end{equation}
where \(\sigma_\perp^2 \equiv \delbx^2 = \delby^2\) is the transverse variance of the magnetic field. Then the mean magnetic field is\footnote{\footnotesize{Note that \(B_0\hat{\bm{z}} \equiv \langle \bm{B}\rangle\), whereas \(\langle B\rangle\equiv \langle|\bm{B}|\rangle\), where \(\bm{B} = b_x\hat{\bm{x}} + b_y \hat{\bm{y}} +(B_0 + b_z) \hat{\bm{z}}\). For purely transverse fluctuations \(b_z=0\).}} 
\begin{equation}
    \langle B\rangle = \int_{B_0}^\infty \frac{B^2}{\sigma_\perp^2} e^{\frac{B_0^2-B^2}{2\sigma_\perp^2}} dB = C_1 I_1. \label{eq:app_mean_B}
\end{equation}
Here \(C_1 = \exp{\left(\frac{B_0^2}{2\sigma_\perp^2}\right)}/\sigma_\perp^2\) and \(I_1 = \int_{B_0}^\infty B^2 \exp{\left(-\frac{B^2}{2\sigma_\perp^2}\right)} dB\). \(I_1\) is an integral of the form  
\begin{equation}
  I = \int_{B_0}^\infty u^2 e^{-\frac{u^2}{a}} du,
\end{equation}
where \(a = 2\sigma_\perp^2\) and \(u=B\). Letting \(t = u^2/a\), we have \(2u du = a dt\), so
\begin{equation}
    I = \frac{a^{3/2}}{2} \int_{t_0}^\infty t^{3/2-1} e^{-t} dt = \frac{a^{3/2}}{2} \Gamma(3/2,t_0),
\end{equation}
where \(t_0 = B_0^2/a\), and we have written the integral as the upper incomplete gamma function \(\Gamma(s,x) = \int_x^\infty t^{s-1} e^{-t}dt\), with \(s=3/2\) and \(x=t_0\) \citep{NIST_DLMF}. Making use of the recurrence relation \(\Gamma(s+1,x) = s\Gamma(s,x) + x^s e^{-x}\), we have 
\(\Gamma(3/2,t_0) = 1/2~\Gamma(1/2,t_0) + t_0^{1/2}e^{-t_0}\). Using the property \(\Gamma(1/2,x) = \pi^{1/2}\text{erfc}(x^{1/2})\), we get 
\begin{equation}
    I = \frac{a^{3/2}}{2}\left[\frac{\sqrt{\pi}}{2} \text{erfc}(t_0^{1/2}) + t_0^{1/2} e^{-t_0}  \right],
\end{equation}
where erfc is the complementary error function. Returning to Equation \eqref{eq:app_mean_B}, we get, after some straightforward algebra,
\begin{equation}
    \langle B\rangle = \sigma_\perp 
    e^{\frac{B_0^2}{2\sigma_\perp^2}}
    \sqrt{\frac{\pi}{2}}
    \text{erfc}\left(\frac{B_0}{\sqrt{2}\sigma_\perp} \right) + B_0.
    \label{eq:app_meanB2}
\end{equation}
For large \(B_0/\sigma_\perp\), \(\langle B\rangle \to B_0\), as expected. 

From Equation \eqref{eq:app_fl3} we have, for purely transverse fluctuations,
\begin{equation}
    \langle S\rangle = \int \frac{\langle B\rangle}{B_0} dz,
    \label{eq:app_fl4}
\end{equation}
which may be integrated along the unperturbed, large-scale field line, together with Equation \eqref{eq:app_meanB2}, to obtain the average path length of random-walking field lines. With the assumption of constant \(\delta b/B_0 = 1\) (or \(\sigma_\perp/B_0 = 1/\sqrt{2}\))
along the field line, 
the above integral gives \(\langle S\rangle = 1.38 S_0\), where \(S_0\) is the path length of the central large-scale field line. This estimate is close to the one obtained in Section \ref{sec:formulation1} from Equation \eqref{eq:ratio_trans}: \(\langle S\rangle \sim 1.41 S_0\). 

\begin{figure}
\centering
\includegraphics[scale=.55]{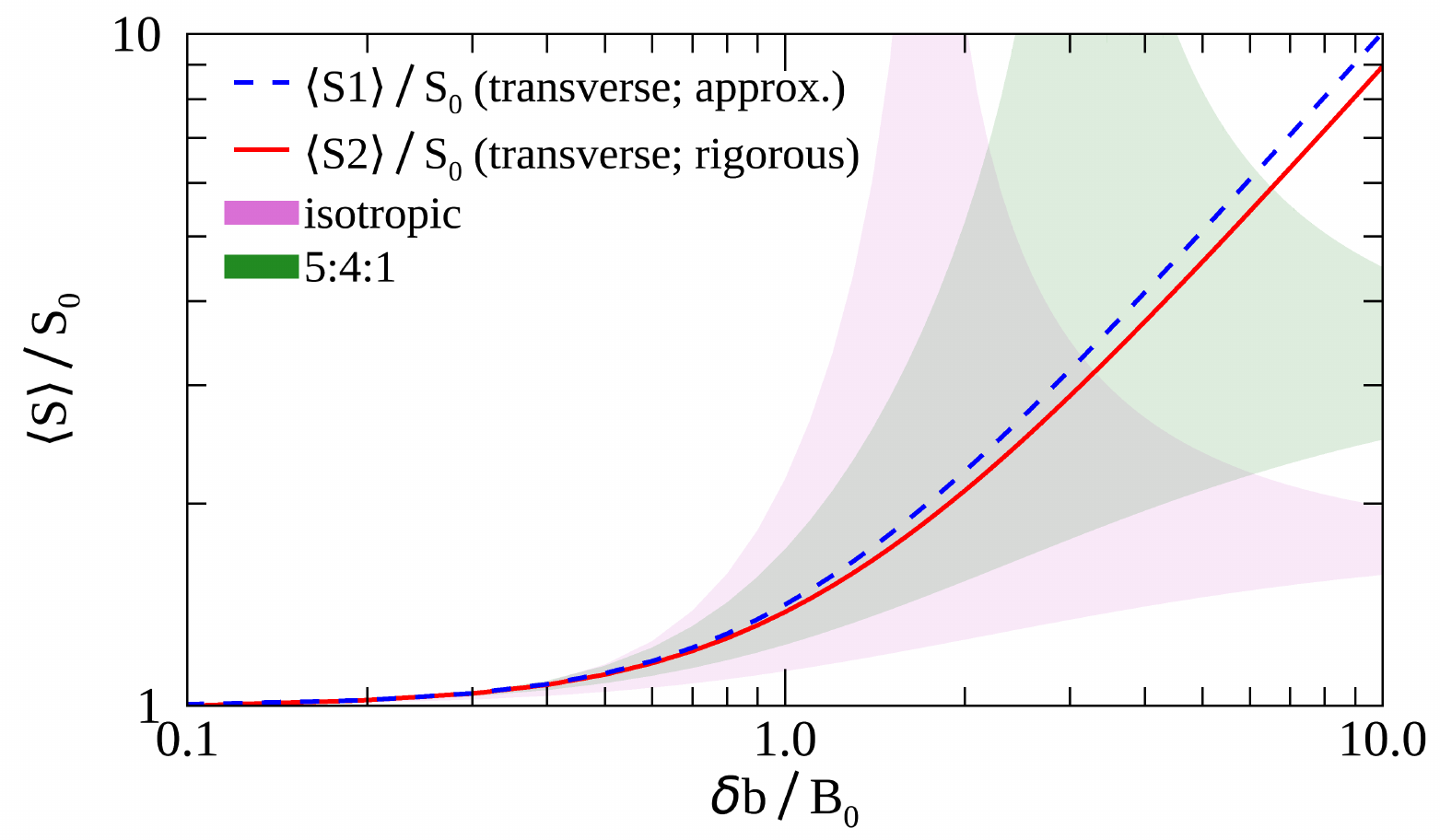}
\caption{Ratio of path length \(\langle S\rangle\) of random-walking field lines to path length \(S_0\) of unperturbed field line, as a function of $\delta b/B_0$. \(\langle S1\rangle\) is based on the simple estimate in Section \ref{sec:formulation1}, while \(\langle S2\rangle\) is based on the more rigorous formalism developed in Section \ref{sec:formulation2}; both these cases are for transverse fluctuations. The pink and green shaded regions represent the cases of isotropic [Equation \eqref{eq:ratio_iso1}] and 5:4:1 [Equation \eqref{eq:ratio_541}] fluctuations, respectively. The lower and upper bounds of the shaded regions correspond to the cases of positive and negative \(\delbz\), respectively (see Section \ref{sec:formulation1}).}
\label{fig:pathlen_vs_ratio}
\end{figure}

Figure \ref{fig:pathlen_vs_ratio} shows four estimates of the path length as a function of the ratio \(\delta b/B_0\). The shaded regions represent the crude estimates for isotropic and 5:4:1 turbulence derived in Section \ref{sec:formulation1}, with the lower and upper bounds corresponding to the cases of positive and negative \(\delbz\), respectively. Note that the upper bounds have a singularity when the denominator of the fractions in Equations \eqref{eq:ratio_iso1} and \eqref{eq:ratio_541} vanishes. The figure also compares the two path length estimates for transverse fluctuations; here \(\langle S1\rangle\) is the simple estimate based on Equation \eqref{eq:ratio_trans}, and \(\langle S2\rangle\) is the rigorous estimate based on Equation \eqref{eq:app_meanB2}. We find that these two estimates are extremely close to each other for \(\delta b/B_0 < 1\); as \(\delta b/B_0\) increases, \(\langle S1\rangle\) becomes slightly larger than \(\langle S2\rangle\). Further, we note that, for all four cases, the increase in path length due to FLRW is fractionally small for \(\delta b/B_0 \lesssim 0.5\). For \(\delta b/B_0 > 2\) the path length can be several times larger than the unperturbed path length, for the three non-isotropic cases.\footnote{\footnotesize{Recall that the isotropic case is not generally of relevance to the solar wind, but could have implications for astrophysical systems \cite[][and references therein]{subedi2017ApJ}. Note also that the turbulence is not generally isotropic unless \(\delta b/B_0 \gg 1\), and therefore the large path lengths seen for the isotropic case in Figure \ref{fig:pathlen_vs_ratio} may not be physically relevant, in the context of the crude model in Section \ref{sec:formulation1}.}} These results suggest that path lengths inferred from SEP observations (see Section \ref{sec:results2}) can potentially provide a measure of the prevailing levels of magnetic fluctuations.

In Section \ref{sec:results1}, we will evaluate \(\langle S\rangle\) along a central field line obtained from a global MHD model of the solar wind, taking into account the spatial variation in \(B_0\) and \(\sigma_\perp\) along the field line.

%
%

%
%

\section{Comparison with Monte Carlo Simulations in a Model of Solar Wind Magnetic Turbulence}

\subsection{Field Line Path Lengths}

Next, we compare the rigorous theoretical result for Gaussian fluctuations from Section 3.2 with the path lengths of field lines as traced by Monte Carlo (MC) simulation based on a model of solar wind magnetic turbulence.
This model, previously described by \cite{ruffolo2013ApJ} and \cite{tooprakai2016ApJ}, uses a superposition of representations of two turbulence components -- a two-dimensional (2D) magnetohydrodynamic (MHD) component and a slab component \citep{bieber1994proton,seripienlert2010apj} -- with a radial mean field of strength $B_0\propto r^{-2}$ in spherical geometry.
The fluctuation amplitude is taken to be proportional to $B_0$, with 20\% of the fluctuation energy in the slab component and 80\% in the 2D MHD component.
The only change we make to the model is to consider different values of the rms magnetic fluctuation amplitude $\delta b$, setting $\delta b/B_0=0.5$ (as in previous work) 
or $\delta b/B_0=1$, to reflect the strong level of magnetic fluctuation observed near the Sun by PSP \citep{bale2019nat}. 

Starting at $r_0=0.1$ AU, we traced 50,000 magnetic field lines from random heliolongitudes and heliolatitudes within a circle of angular radius 2.5$^\circ$ and measured their incremental path length $\Delta S$ over a distance $\Delta r=0.25$ AU to $r=0.35$ AU, close to the radius of PSP observations considered in this work.

\begin{figure}
\centering
\includegraphics[scale=.45]{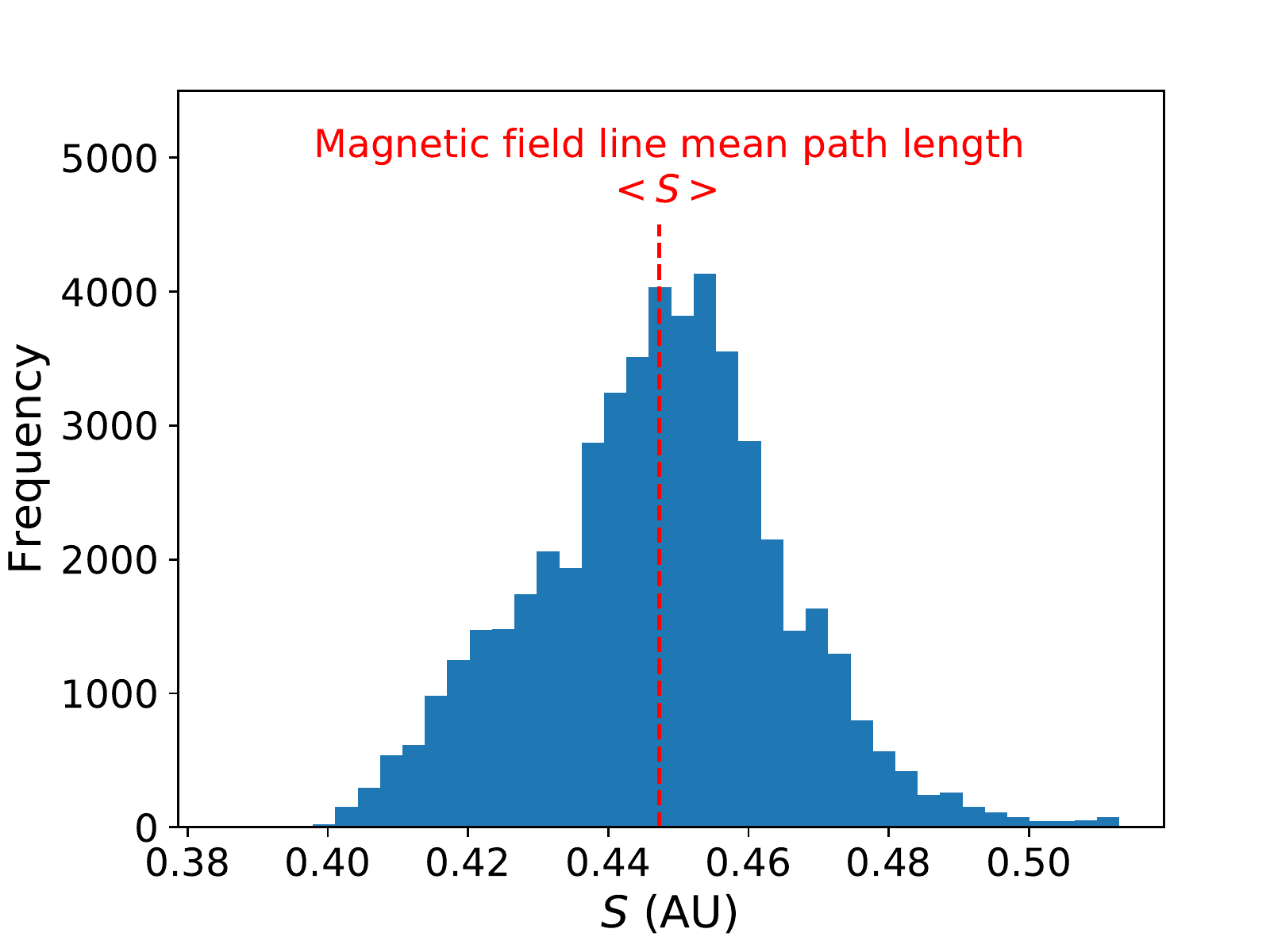}
\caption{Distribution of path length $S$ of magnetic field lines computationally traced 
to $r=0.35$ AU
in a model of solar wind turbulence (a representation of 2D MHD + slab magnetic turbulence in spherical geometry), superposed on a radial mean field of strength $B_0\propto r^{-2}$. 
The rms turbulent amplitude was set equal to the mean field, and the slab energy fraction to 0.2.  
Vertical dashed line indicates the mean value $\langle S\rangle=0.447$ AU, 
which is close to our theoretical result of 0.445 AU.}
\label{fig:fl1}
\end{figure}
\begin{figure}
\centering
\includegraphics[scale=.6]{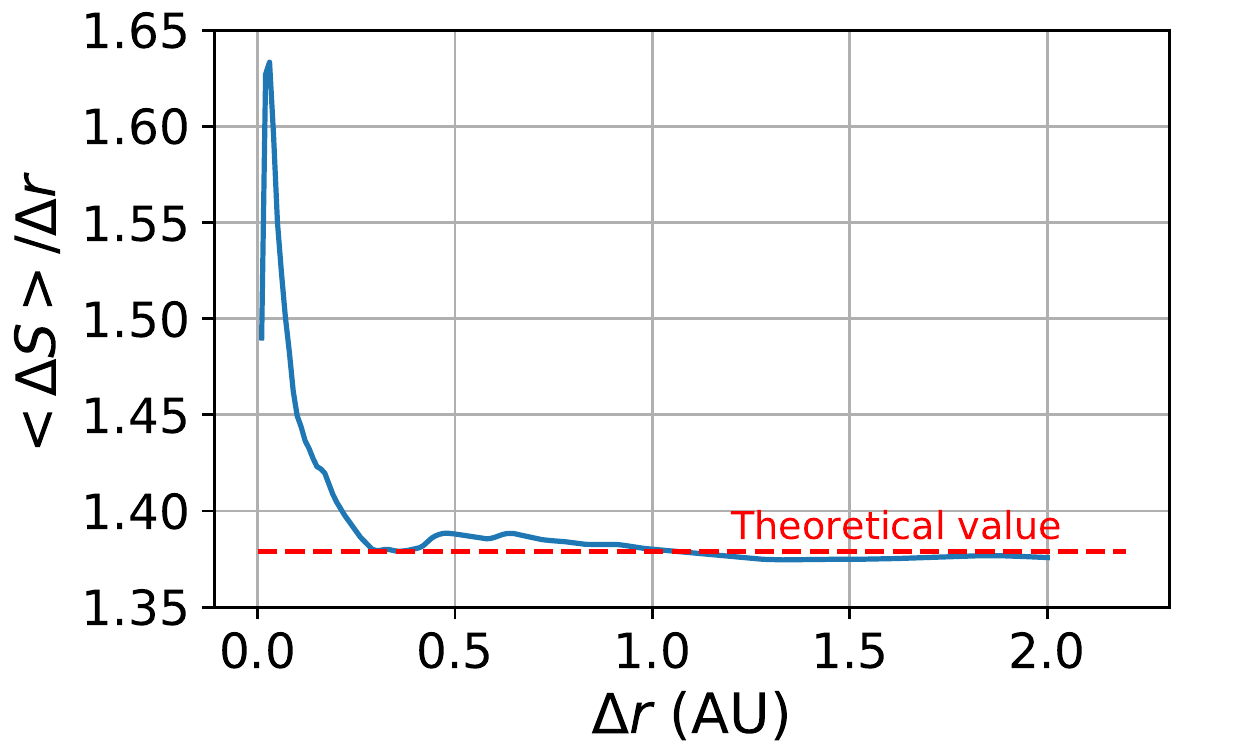}
\caption{Ratio of average incremental magnetic field line path length $\langle \Delta S\rangle$ to radial distance $\Delta r$ as a function of $\Delta r$, for computations as described for Figure \ref{fig:fl1}, in comparison with the theoretical ratio of 1.379.}
\label{fig:fl2}
\end{figure}

Since the model uses a constant $\delta b/B_0$, even though $\delta b$ and $B_0$ individually vary as $r^{-2}$, we expect that the average incremental path length of magnetic field lines over $\Delta r$ should be 
\begin{eqnarray}
\frac{\langle \Delta S\rangle}{\Delta r} &=& \frac{\langle B\rangle}{B_0} \\
&\approx& 
1+\frac{\sqrt{\pi}}{2}\frac{\delta b}{B_0} 
    \exp\left(\frac{B_0^2}{\delta b^2}\right)
    \text{erfc}\left(\frac{B_0}{\delta b} \right).\label{eq:meanS}
\end{eqnarray}
Here we have made use of  Equation (\ref{eq:app_meanB2}) from the rigorous theory for Guassian fluctuations.
Actually the 2D MHD field that we use does not have a Gaussian distribution of transverse components.
The kurtosis of $\approx$2.7 \citep{seripienlert2010apj} indicates a moderate departure from the Gaussian value of 3, because MHD tends to make the magnetic pressure and magnetic field magnitude more uniform over small scales.
Nevertheless, we find that the theory for Gaussian fluctuations from Section 3.2 provides a good match to our simulation results.

We also consider a total pathlength $S=\Delta S+r_0$ to account for the field line distance between the Sun and $r_0$.
Adding the constant $r_0$ can be justified to some degree by noting that the Alfv\'en critical zone, where the solar wind speed roughly equals the Alfv\'en speed, may be near $r\sim0.1$ AU.  
The solar wind turbulent energy is expected to peak in this region and to be weaker at lower $r$, where coronal flux tubes may be more rigid \citep{chhiber2019psp1,ruffolo2020ApJ}. Consequently, in this implementation, and in general, 
the path length increase due to field line random walk is expected to be small in the sub-Alfv\'enic inner corona.

Figure \ref{fig:fl1} shows the simulated distribution of magnetic path length $S$ 
traced to $r=0.35$ AU
for the case of $\delta b/B_0=1$.  The average simulated path length $\langle S\rangle$ is 0.447 AU (vertical dashed line), representing a 28\% increase over the distance parallel to the large-scale field.  From the theory, we have $\langle B\rangle/B_0=1.379$, which by Equation (\ref{eq:meanS}) implies $\langle \Delta S\rangle\approx0.345$ AU and $\langle S\rangle\approx0.445$ AU.
This provides a close match to our simulation result.
For $\delta b/B_0=0.5$ we also find good agreement, for a simulation result of $\langle S\rangle=0.382$ AU (a 9\% increase over the parallel distance) and theory result of $\langle B\rangle/B_0=1.113$ and $\langle S\rangle=0.378$ AU.

For $\delta b/B_0=1$, Figure \ref{fig:fl2} shows the enhancement of the incremental field line path length $\langle\Delta S\rangle$ relative to $\Delta r$, the traced distance along the large-scale magnetic field, as a function of $\Delta r$.  There is an excess in the average field line path length over the theoretical value for short distances.  When tracing over $\Delta r=0.25$ AU or longer, the simulation results remain close to the theoretical value.

It is interesting that the theory matches these simulation results well, despite moderate departures from Gaussianity in the dominant magnetic fluctuation component. This agreement gives us greater confidence in applying the theory to address observations of SEP transport in the actual solar wind.

\subsection{Particle Guiding-center and Full-orbit Path Lengths}\label{sec:MC2} 

We also performed full-orbit trajectory tracing of 50,000 protons in the same representation of 2D MHD + slab magnetic fluctuations superposed on a radial magnetic field.  
We measured both the path length $s$ along the full orbit (including the particle gyromotion) and the path length $s_c$ of the guiding center.  
The guiding center location was calculated from the instantaneous particle position \(\bm{r}\) and momentum \(\bm{p}\) from the full orbit particle tracing by
\begin{equation}
    \bm{r}_\text{GC} = \bm{r}-\frac{\bm{B}\times{\bm{p}}}{qB^2}.
\end{equation}

We recorded the values of $s$ and $s_c$ for particles whenever they crossed the radius of interest, in this case $r=0.35$ AU.  
Multiple crossings are included, to allow for backscattering of particles from higher $r$, as such particles are included in actual SEP observations. 
Because magnetostatic fluctuations do no work on a particle, the speed (magnitude) $v$ is a constant of the motion.  
Therefore, the orbit path length beyond $r_0$ can be calculated simply as $vt$, where $t$ is the time of arrival of the particle at the radius of interest relative to its release from $r=r_0$.
The guiding center path length is calculated by summing $\Delta s_c$ from each time step in the simulation. 
Our simulations start tracing field lines and particles at $r_0=0.1$ AU; to account for this and facilitate comparison with SEP observations, we define the total path lengths as $s=vt+r_0$ and $s_c=\sum\Delta s_c+r_0$, and also add $r_0$ to the magnetic field line path lengths for comparison.
The same definition of $s$ was employed by \citet{tooprakai2016ApJ}.
This effectively assumes that field lines had negligible fluctuation at $r<r_0$ and that particles traveled along field lines with zero pitch angle;  
strong adiabatic focusing (magnetic mirroring) near the Sun tends to make the pitch angle distribution concentrated near zero within $r=0.1$ AU \citep{ruffolo1995GRL}, becoming less concentrated thereafter due to pitch angle scattering.

\begin{figure}
\centering
\includegraphics[scale=.55]{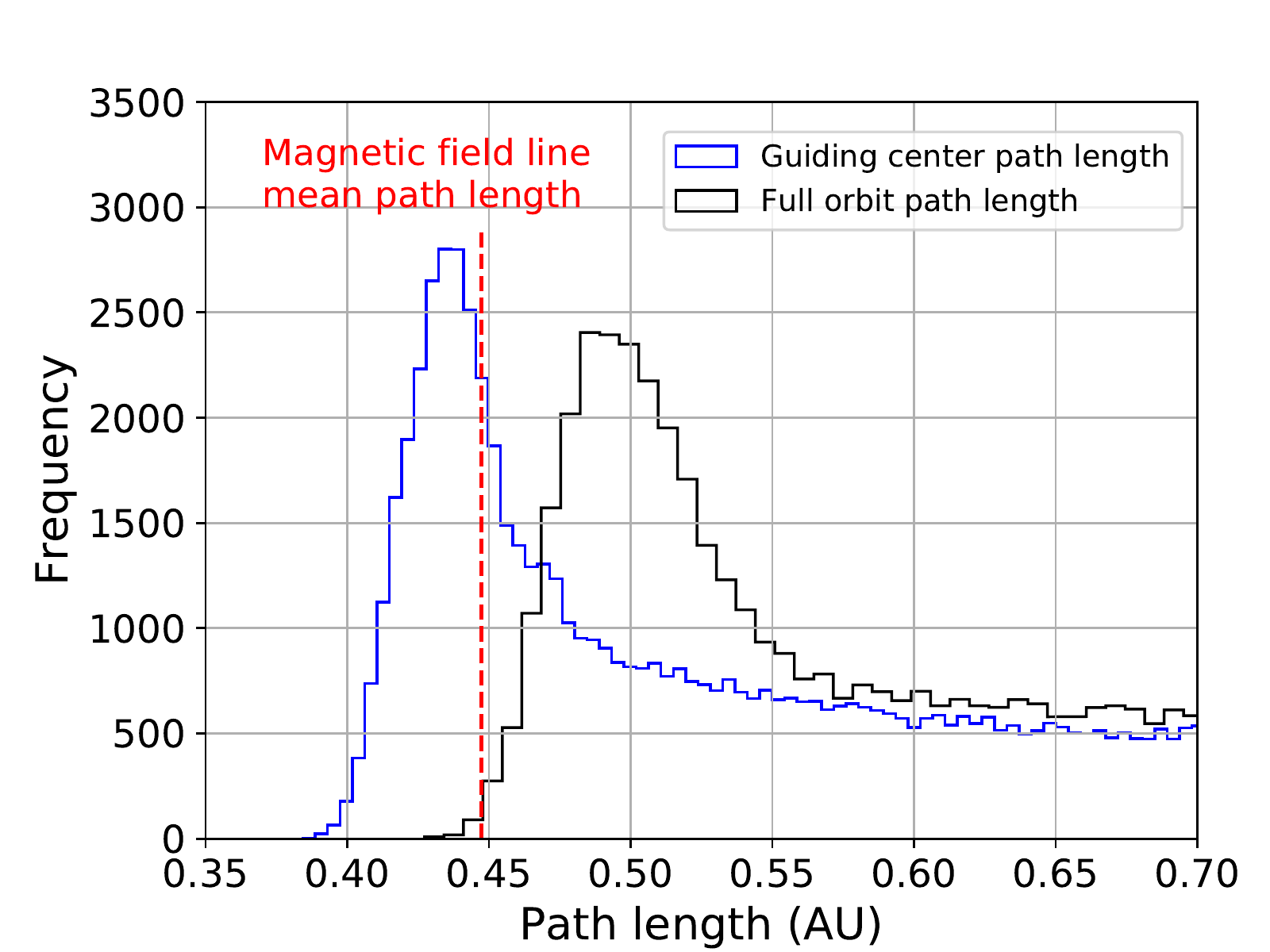}
\caption{Histograms of simulated path lengths of guiding center motion and full orbital motion for 25 MeV protons arriving promptly at radius $r=0.35$ AU from the Sun, for computations as described for Figure \ref{fig:fl1} and in the text, in comparison with the average path length of magnetic field lines.
The guiding center path length of promptly arriving particles (before the arrival of the backscattered population) is usually shorter than the average path length of magnetic field lines because of the finite radius of the gyromotion, which implies that the particle samples the magnetic field over a range of positions and need not exactly follow the random walk of individual field lines.
The full orbit path length of the particles is longer because the gyromotion implies that particles follow a longer path than their guiding centers.
}
\label{fig:gc+orbit}
\end{figure}

The results for a proton kinetic energy of 25 MeV and $\delta b/B_0=1$ are shown in Figure \ref{fig:gc+orbit}.
Note that distribution of the orbit path length $s=vt+r_0$ (black) directly corresponds to the time-intensity profile of SEP observation.
The distribution of the guiding center path length $s_c$ corresponds to the time-intensity profile that would be observed if there were no gyromotion, i.e., for guiding center transport at pitch angle 0$^\circ$ or 180$^\circ$.
Because of backscattering, both distributions have a ``wake'' that extends to indefinitely long pathlength \citep[or arrival time;][]{earl1976b}.  
Therefore, we do not use mean values of $s_c$ and $s$ to characterize the distributions, and instead consider the peak path lengths and minimum path lengths.

It is interesting to check whether the particle guiding center actually follows a magnetic field line, with the same path length.
One might imagine that a particle, with its finite radius of gyration, averages over fluctuations on scales smaller than that radius, and its guiding center might have a shorter path length than the field lines.
Indeed, Figure \ref{fig:gc+orbit} demonstrates this effect for the case of $\delta b/B_0=1$, showing that the peak guiding center path length of promptly arriving particles, at $\approx0.435$ AU, is slightly shorter than the average magnetic field line path length 
$\langle S\rangle=0.447$ AU for the same simulation.
In the case of weaker turbulence amplitude, $\delta b/B_0=0.5$, our simulation results are consistent with no difference between the peak guiding center path length and the average field line path length (both close to 0.38 AU), which in this case are only $\approx10$\% longer than the radial distance.


As seen in Figure \ref{fig:gc+orbit}, the full orbit path length $s$ is distributed over longer values than the guiding center path length $s_c$.
In general, this must be the case whenever the pitch angle is nonzero; locally $ds_c=|\mu| ds$ where $\mu$ is the pitch angle cosine.
Here we find that the nonzero pitch angle and gyromotion of the particles leads to a substantial increase in the particle path length expected for PSP observations, as previously predicted for observations near 1 AU \citep{lintunen2004AA,saiz2005ApJ}.

Note that the minimum full orbit path lengths were presumably associated with near-minimal guiding center path lengths, and the minimum full orbit path lengths are substantially longer.
Therefore, even the first arriving particles underwent transport characterized by nonzero pitch angle.
In fact, we can make use of the relation $ds_c=|\mu| ds$, assuming that $\mu>0$, to define an effective pitch angle from $\mu_{\rm eff}=s_c/s$.  
For the case shown in Figure \ref{fig:gc+orbit}, the ratio of either minimum values of $s_c$ and $s$ or peak values of these quantities yields essentially the same value of $\mu_{\rm eff}=0.90$-0.91, corresponding to an effective pitch angle $\theta_{\rm eff}\approx25^\circ$.
We have also verified that the distribution of $\mu_{\rm eff}$ for individual particles, grouped by their orbital path length $s$, contains no ``scatter free'' particles with $\mu_{\rm eff}=1$; rather, the distribution is clustered around a mean value that is consistent with above ratio and has very little change from event onset to peak.

Note that estimates of the SEP path length from observations are typically based on the first detected particles.  
It is interesting that for the case shown in Figure \ref{fig:gc+orbit}, the average path length of magnetic field lines provides a good indicator of the minimum orbit path length $s$ and time of arrival $t$ of the first particles.
This may be somewhat of a coincidence based on competing effects.
The minimum orbit path length of observed particles may be associated with the minimum guiding center pathlength, which is shorter than the average guiding center path length, which is in turn shorter than the average magnetic field path length as described above.  
But the increase of the minimum orbit path length over the minimum guiding center path length, due to gyromotion, mostly compensates in this case to bring the minimum orbit path length close to the average field line path length.

In summary, for $\delta b/B_0=1$ and protons of $E=25$ MeV, we find that compared with the total radial distance of 0.35 AU, the average simulated magnetic field line pathlength is longer by 0.097 AU (due to the field line random walk), the peak guiding center path length is shorter than that by about 0.012 AU (due to the gyromotion averaging over fluctuations to some degree), and the peak full orbit path length is longer than that by 0.045 AU (due to the gyromotion itself).  Thus in this case the increase in path length is mainly associated with the magnetic field line random walk.

For $\delta b/B_0=0.5$ at the same particle energy, the average simulated magnetic field path length is longer than the radial distance by only 0.032, the peak guiding center path length ($\approx$0.38 AU) is about the same, and the peak full orbit path length ($\approx$0.42 AU) is longer by 0.04 AU.
In this case of weaker turbulence, the increase in path length can be attributed nearly equally to the field line random walk and the gyromotion.

\begin{figure*}[h]
\centering
\includegraphics[scale=.5]{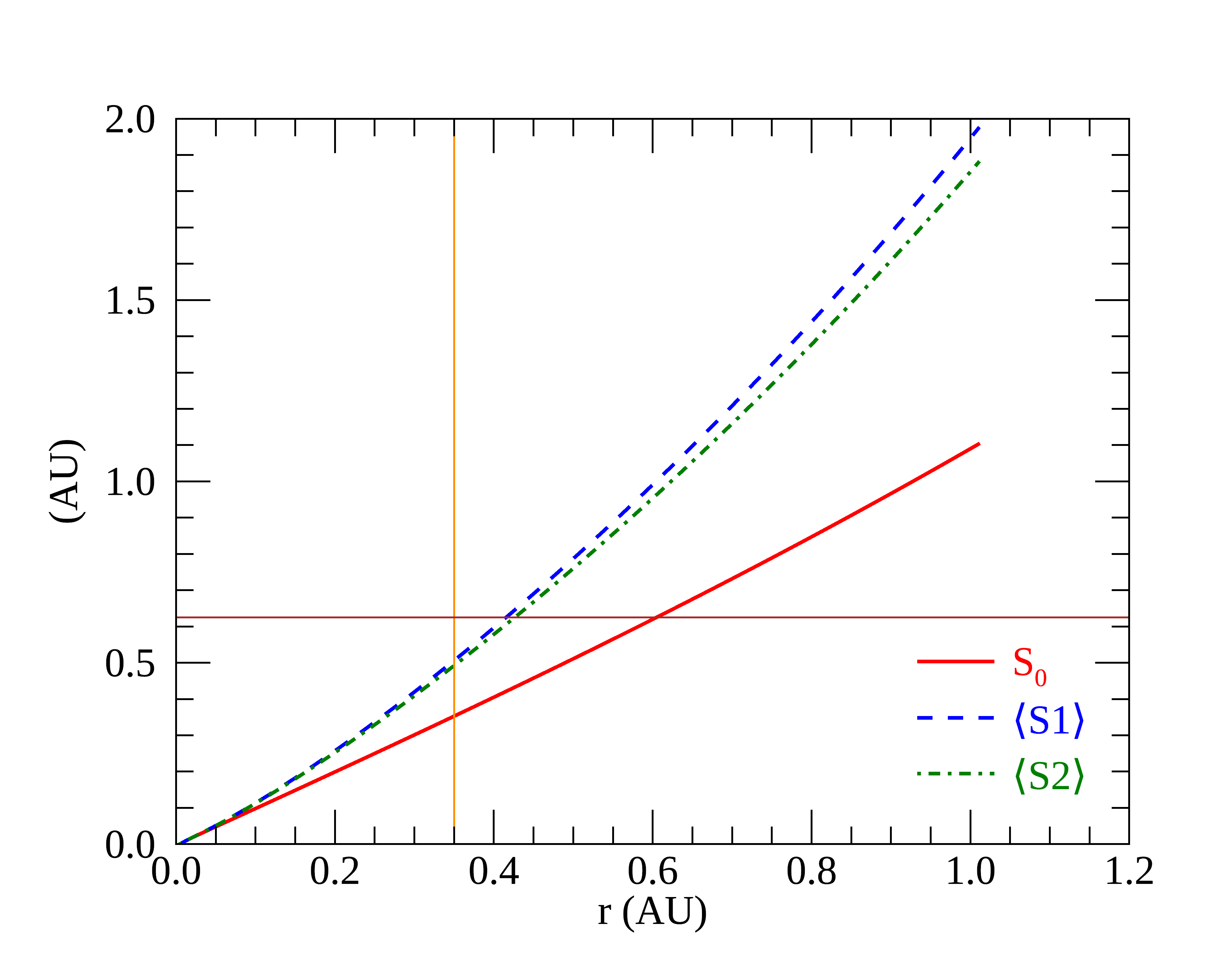}
\includegraphics[scale=.5]{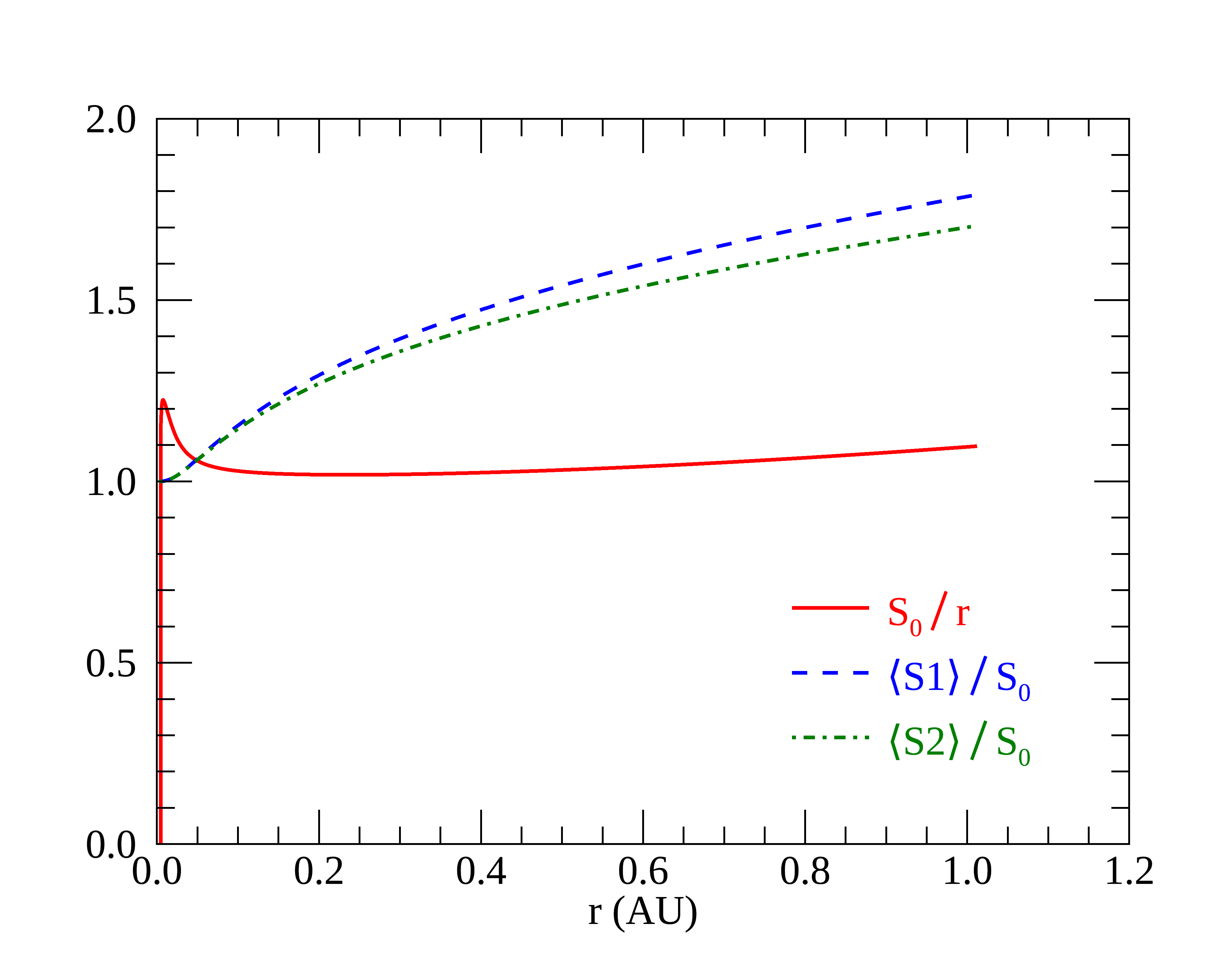}
\caption{Path length \(S_0\) vs heliocentric distance \(r\) for a selected large-scale field line from a global heliospheric simulation based on a solar magnetogram for 28 May 2020, compared with two computations of the average path length of random-walking field lines associated with that particular large-scale field line. \(\langle S1\rangle\) is based on the simple estimate in Section \ref{sec:formulation1}, while \(\langle S2\rangle\) is based on the more rigorous formalism developed in Section \ref{sec:formulation2} (see text). Both cases are for transverse fluctuations. In the left panel, the orange vertical line marks the location of PSP at the time of observation of the energetic ion events discussed in Section \ref{sec:results2}, and the brown horizontal line marks the particle path length inferred from these observations. See also Figure \ref{fig:pathlens}.}
\label{fig:sim2_pathlen}
\end{figure*}

\section{Application of Theory to Solar Energetic Particle Transport in the Solar Wind}

\subsection{Field Line Path Length in Global Heliospheric Simulation}\label{sec:results1}
The Usmanov global heliospheric MHD simulation model \citep{usmanov2014three,usmanov2018} solves compressible three dimensional MHD equations for mean, or large-scale, MHD variables, and incorporates a turbulence transport model that self-consistently interacts with the resolved simulation variables. This code accounts well for large-scale features of the interplanetary medium as observed by Ulysses and Voyager \citep{usmanov2012three,usmanov2018}, as well as turbulence properties observed by PSP (Chhiber et al. 2021a; in prep). The code has been used to evaluate energetic particle diffusion coefficients throughout the heliosphere \citep{chhiber2017ApJS230}, and to provide several types of contextual predictions for PSP \citep{chhiber2019psp1,chhiber2019psp2}. 
Because this model provides dynamical
solutions for the large-scale magnetic field as well as the rms turbulence amplitude, it can provide all information necessary to evaluate the magnetic field path lengths using the formulation given in the previous section. Here we use a simulation based on an 
ADAPT solar magnetogram \citep{arge2010AIPC} corresponding to 2020 May 28 -- the time of the PSP observations examined here. The fluctuations in the turbulence transport model are purely transverse relative to the mean field, and an Alfv\'en ratio of 0.5 is assumed. For more details on the simulation, including boundary and initial conditions, see \cite{usmanov2014three} and \cite{usmanov2018}.

Figure \ref{fig:sim2_pathlen} shows the path length \(S_0\) for a selected large-scale field line, compared with two computations of the average path-length of random-walking field lines associated with that particular large-scale field line: the simple estimate \(\langle S1\rangle\) is computed by integrating Equation \eqref{eq:app_fl3}, using Equation \eqref{eq:ratio_trans}, while the rigorous estimate \(\langle S2\rangle\) is computed using Equation \eqref{eq:app_fl4} with \(\langle B\rangle\) specified by Equation \eqref{eq:app_meanB2}. Note that both \(\delta b\) and \(B_0\) vary along the field line. 

We find that the Parker-spiral-like path length \(S_0\) is \(\sim 1.1\) AU at a heliocentric radius of 1 AU, while the path length of random-walking field lines is nearly 2 AU at that distance. Clearly, the FLRW can produce a significant increase in path length of magnetic field lines, relative to the unperturbed field line. We also note that, for \(r \gtrsim 0.4\) AU, \(\langle S1\rangle\) becomes noticeably larger than \(\langle S2\rangle\), due to a slight and gradual increase in the ratio \(\delta b/B_0\) with heliocentric distance (see also Figure \ref{fig:pathlen_vs_ratio}). The PSP observations annotated in the left panel of Figure \ref{fig:sim2_pathlen} will be discussed in Section \ref{sec:results2}, below.

\subsection{Application to Parker Solar Probe/\isois~SEP events}\label{sec:results2}

\begin{figure}
\centering
\includegraphics[scale=.33]{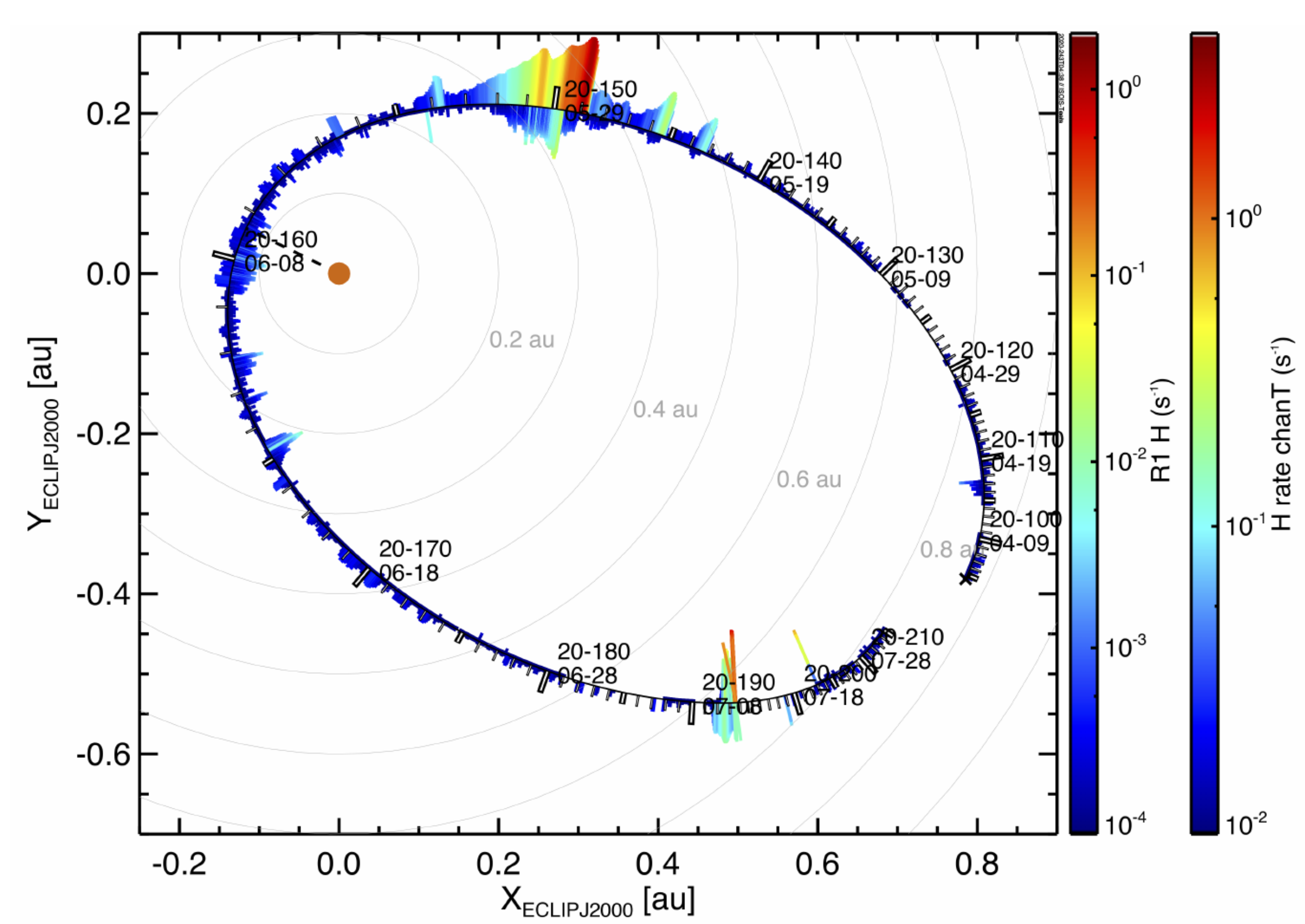}
\caption{PSP Orbit 5 highlighting SEP events observed by \isois~during 2020 April-July.
Count rate is indicated by both the color and the length of the bars, for particles (primarily H+) at lower (about 30–
200 keV; inside track; right color scale) and higher energies (about 1–2 MeV/nuc; outside track; left color scale). 
Intervals without data are indicated by the black orbital track.  
The time scale is indicated by ticks and white rectangles on the outer track as YY-DOY and MM-DD, where DOY is day of year (UT).
}
\label{fig:orbit0}
\end{figure}

The Parker Solar Probe mission is currently on its seventh orbit of the Sun, progressively descending
to perihelia deeper in the solar corona with each swing by Venus \citep{fox2016SSR}. 
Energetic particle (EP) data from the \isois~suite cover a wide range of energies using two instruments \citep{mccomas2016SSR}.
EPI-Lo measures energetic ions from 0.02 to \(\sim 1.5\) MeV/nuc and energetic electrons of 25–1000 keV. 
The EPi-Hi instrument 
measures energetic protons and He nuclei from \(\sim 1\) to \(\sim 100\) MeV/nuc
(and higher energies for heavier elements) and energetic electrons from \(\sim 0.5\) to \(\sim 6\) MeV.
To cover this energy range, and to provide wide FOV coverage, EPI-Hi has three telescopes,
a double-ended high energy telescope (HET), 
with apertures HETA and HETB, a double-ended low energy telescope (LET1), with apertures LETA and LETB, and a single-ended low energy telescope (LET2).

\begin{figure*}
\centering
\includegraphics[scale=.65]{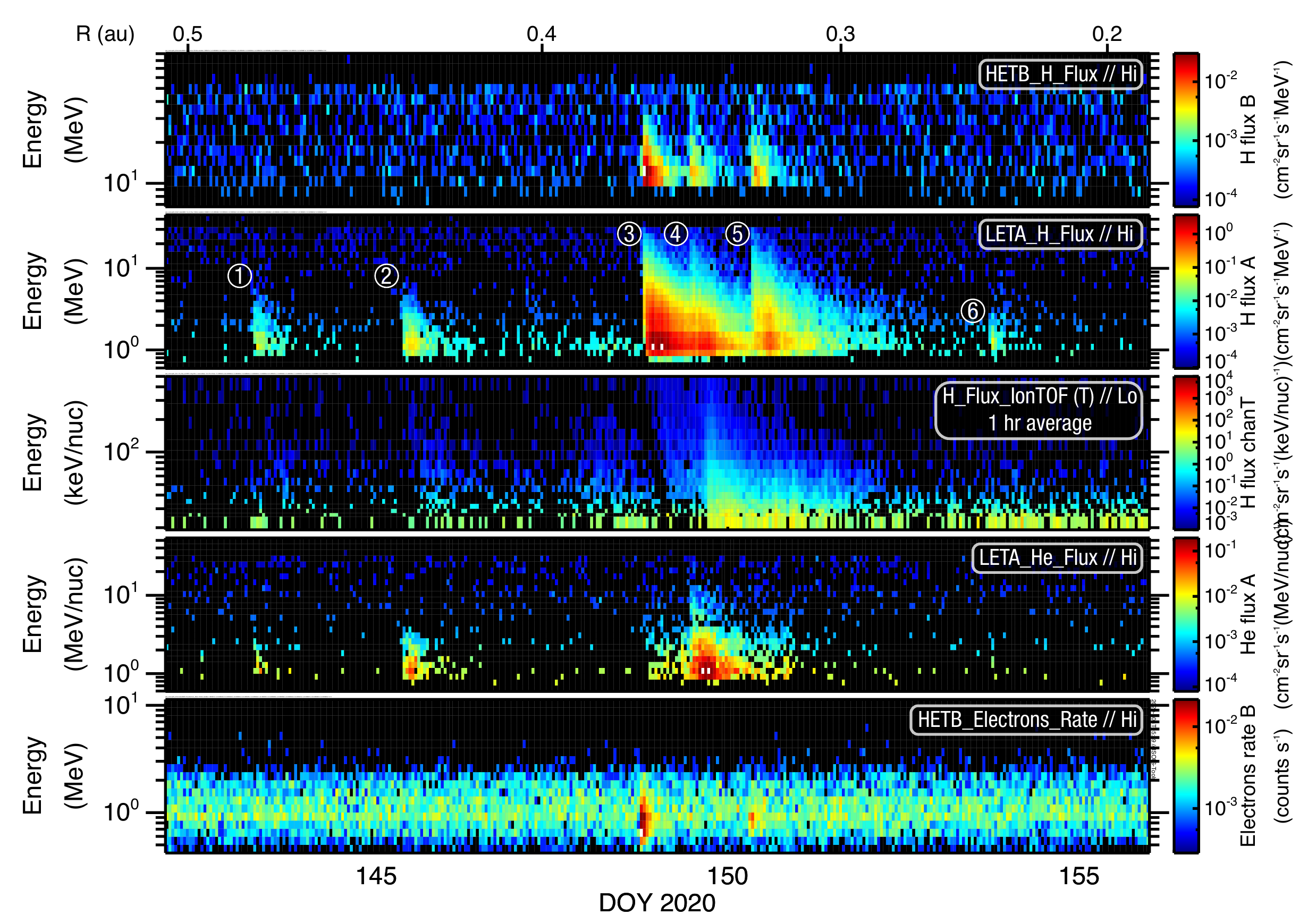}
\caption{SEP Events observed by \isois\ from 2020 May 21 (DOY 142) to June 3 (DOY 155), near 0.35 AU.
Intensities are indicated by the color scale, as a function of kinetic energy per nucleon and time, for
protons (from HETA, top panel; LETA,
second panel; and EPI-Lo, third panel), helium (from
LETA, fourth panel), and electrons (in counts/sec from HETB, bottom
panel). The six events are most clearly seen in the LETA
proton spectrogram (second panel).
}
\label{fig:orbit1}
\end{figure*}

During Orbit 5, in a several day period from 2020 May 22 to June 2, \isois~measured at least six distinct SEP events. These are depicted in Figures \ref{fig:orbit0} and \ref{fig:orbit1}. The orbit and the general features of the energetic particle fluxes measured by \isois~are illustrated in Figure \ref{fig:orbit0}. The EPi-Hi instrument count rates are indicated by bars of varying sizes on the outside of the orbit; the EPi-Lo count rates are given on the inside of
the orbit curve. Figure \ref{fig:orbit1} shows data from EPI-Hi and EPI-Lo, and gives more detail of the six easily recognizable energetic particle events during the period from 2020 May 22 (day 143) to 2020 June 2 (day 154). They are numbered in time ordering from 1 to 6. During this target period, PSP was at an approximate Sun-centered radial distance of 0.35 AU. These events are discussed in greater detail in Cohen et al. (2021, present issue). We note that for a given particle energy (speed), the intensity vs.\ time in Figure 8 
appears to exhibit a rapid rise to a peak intensity, followed by a more gradual decline, which is qualitatively consistent with the peak and the long ``wake'' at later time (longer path length) for particle orbits from the Monte Carlo simulation (Figure \ref{fig:gc+orbit}).

\begin{figure}
\centering
\includegraphics[scale=.34]{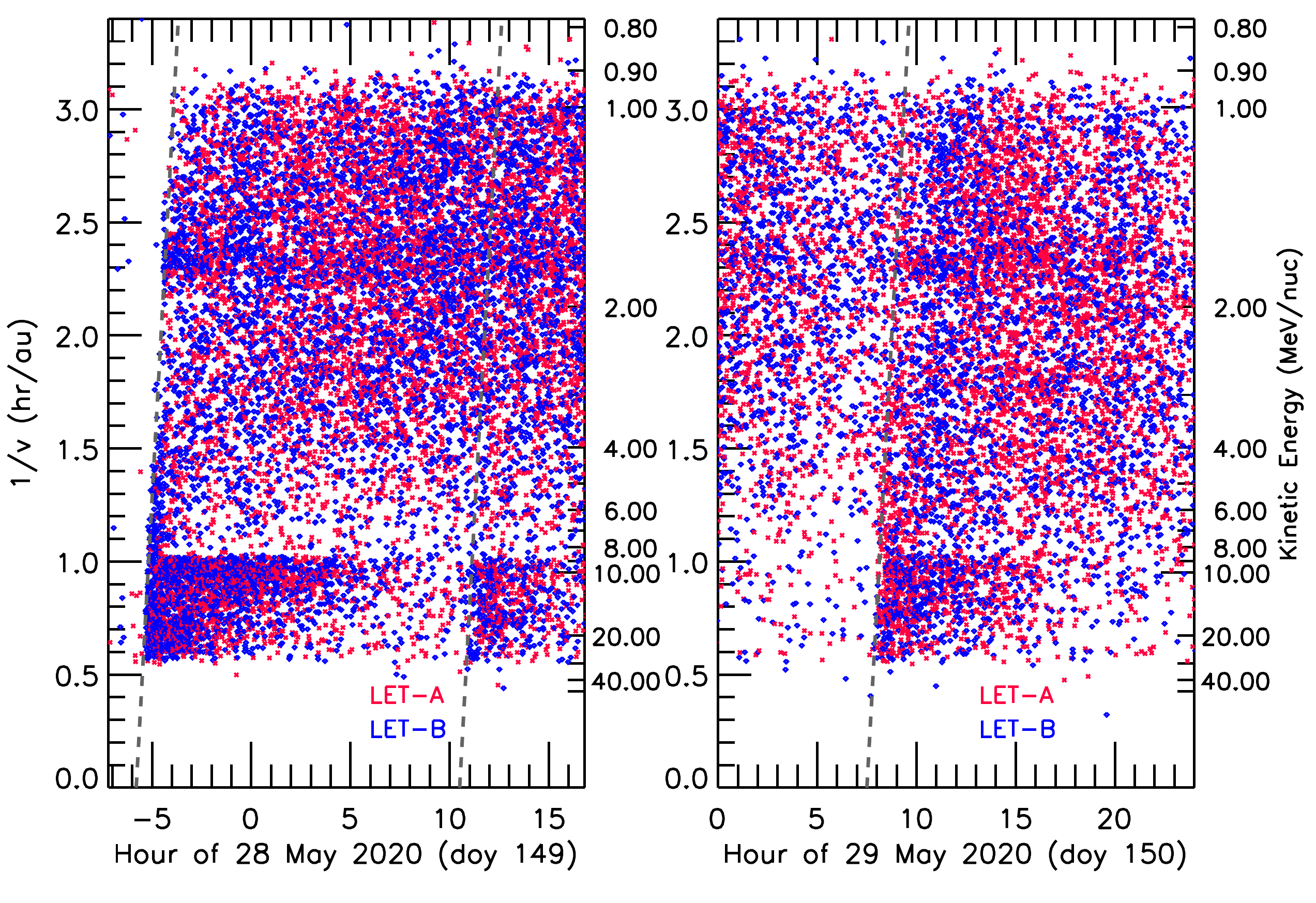}
\caption{Dispersion analysis of $1/v$ vs.\ arrival time for protons
during events 3 \& 4 (left) and event 5 (right). Each point is a measured proton from either LETA (red) or LETB (blue).  The slopes at the onset of each event lead to an estimated path length of about 0.625 AU.}
\label{fig:orbit2}
\end{figure}

To determine an estimated path length for these SEP events, we carried out a standard velocity dispersion analysis, using the data from the LETA and LETB detectors on EPI-Hi. The procedure is to convert the total energy of each measured ion into reciprocal velocity $1/v$ and plot it versus the observation time. In this format the first arriving particles are generally at lower $1/v$  (i.e., higher energy), with lower-speed particles arriving later. A straight line fit to the first arriving particles provides a path length (from the slope) and a release time at the source (from the x-intercept).  The two panels of Figure \ref{fig:orbit2}
show such an analysis on 27-28 May (left) for SEP events 3 \& 4,  and on 29 May (right) for SEP event 5. One readily observes a sharp onset of SEP event 3, starting around five hours prior to 28 May in the left panel. 
Analysis of the slope of line implies 
a path length of approximately 0.625 AU
from the source to the point of observation at PSP.
Transcribing a line with the same slope to the temporal positions of the other events 4 \& 5 as
shown in Figure \ref{fig:orbit2} indicates
that the same slope, and therefore 
the same distance from source to observation, also works well for those cases. This is significantly longer than  PSP's heliocentric distance of about 0.35 AU. Note that a similar result was obtained by \cite{Leske2020ApJS} for the SEP event observed by PSP on 4th April 2019, where the inferred path length was 0.35 AU while the spacecraft was at a heliocentric distance of 0.17 AU.

\begin{figure}
\centering
\includegraphics[scale=.5]{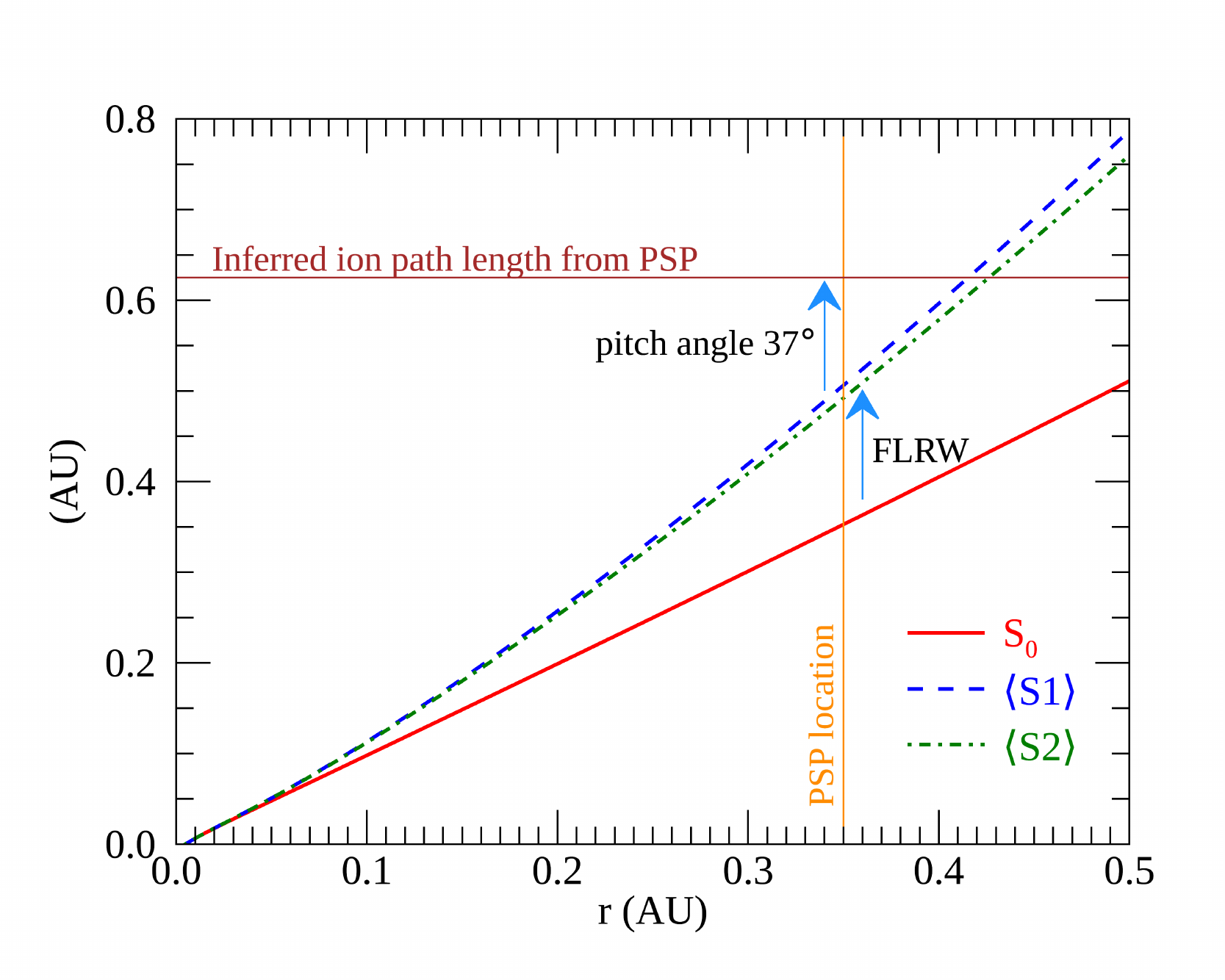}
\caption{A blow up of Figure \ref{fig:sim2_pathlen}, focusing on smaller heliocentric distances. Path length \(S_0\) for a large-scale field line from the global solar wind simulation is compared with two computations of path length of random-walking field lines (see caption of Figure \ref{fig:sim2_pathlen}). Orange vertical line indicates the approximate radial position of PSP at the time of the events depicted in Figure \ref{fig:orbit1}, and brown horizontal line depicts the path length derived from the dispersion analysis of these events, shown in Figure \ref{fig:orbit2}.}
\label{fig:pathlens}
\end{figure}

We may employ the analysis described above, and the estimation of the field line and particle path lengths given in Sections \ref{sec:MC2} and \ref{sec:results1} to offer an explanation of this path length.
In fact three estimates of field
line length can be given as described above (see Figure \ref{fig:sim2_pathlen}). As before, these are designated as $S_0$, $\langle S1 \rangle$, and $\langle S2 \rangle$. These estimates are depicted in Figure \ref{fig:pathlens} for the range of distances encompassing PSP's position at the time of our six targeted events in Figure \ref{fig:orbit2}. We immediately observe that the resolved field line length $S_0$ significantly underestimates the path length derived from the EPI-Hi dispersion analysis in Figure
\ref{fig:orbit2}. However, the two estimates derived using the field line random walk corrections, first, based on the average variance $\langle S1 \rangle$,
and second, based the more complete stochastic theory
$\langle S2 \rangle $, revise the estimated path length, moving it substantially closer to the path length obtained from dispersion analysis.

The remaining deficit  relative to the observation
is presumably accounted for by particle orbit effects. 
In Section \ref{sec:MC2} we quantified the latter
effect for Monte Carlo simulation of a test case, 
finding that the discrepancy between field line path length and particle orbit path can be conveniently parameterized by an effective particle pitch angle of 25$^\circ$. In a similar way, we can estimate an effective pitch angle for the SEP events observed by PSP/\isois\ during 2020 May-June. 
For this estimation we make two assumptions: 
1) Minimum and peak times of particle intensity are very close in time, so we estimate the peak particle orbit path length as 0.625 AU. 
2) We neglect the  difference in path length between the peak guiding center path length and the average magnetic field path length.  
With these assumptions we can estimate that $\cos(\theta_{\rm eff})\approx \langle S2\rangle / (0.625~\text{AU})$, from which we obtain an effective pitch angle of $\theta_{\rm eff}\approx37^\circ$. This is not much greater than the value inferred from the MC simulation, and it is not unreasonable that the SEP transport during these events may have been more diffusive than that in the MC simulation.\footnote{\footnotesize{Note that the MC simulation used a constant \(\delta b/B_0\), while this ratio is spatially varying in the global solar wind simulation.}}

\section{Conclusions}\label{sec:conclude}

Determining the path length for transit of solar energetic particles from source to point of observation is a
subtle and even elusive problem.
Many factors might enter, likely varying from case to case, including free streaming, curvature and distortion of the large-scale magnetic field, topological trapping, parallel scattering, perpendicular diffusion, 
subdiffusion, time dependence of the magnetic field, and field line random walk. In this paper we have chosen to concentrate on two effects, both simplified, namely the effect of average field line length including FLRW, and the effect of 
particle gyromotion on the full
orbit path length,
parameterized by an effective pitch angle. 

The FLRW path length sets a natural scale for the problem, even if some of the other effects are also relevant. The orbital 
path length calculation
incorporates 
a simple treatment 
of the extension 
of path length 
that occurs when particles
follow random-walking field lines, but with nonzero pitch angle, implying a path length systematically longer than the field line length itself. 
 
Our approach 
consisted of analytical estimation and validation with numerical tests.  First, we developed 
an exact analytical theory for the average pathlength using 
the assumption of classical FLRW, a 
Gaussian distribution of magnetic 
fluctuation components, 
and transverse fluctuations.
A simpler estimation based on variances of the magnetic fluctuations is also provided. 
Based on these results
we carried out an example application using a global heliospheric MHD simulation that 
includes self-consistent 
turbulence modeling. 
These two 
FLRW approaches give similar results. These
were validated using  
a Monte Carlo analysis 
of an ensemble of field lines in a synthetic
turbulence model, with 
good results. 
We note that due to the lower 
relative strength of fluctuations compared to the mean magnetic field
in the inner sub-Alfv\'enic corona \citep{chhiber2019psp2},
the contribution to increased path 
length of fieldlines is expected  
to be much lower in that region. 
Based on this reasoning
the FLRW contribution
was ignored in the Monte
Carlo analysis below 0.1 AU, 
and this is  
corroborated by the 
results based on global simulation
shown in Figure \ref{fig:sim2_pathlen}.

To go beyond the field line path length estimation, we considered particle orbit effects.  To this end, 
a second Monte Carlo analysis followed test particles (protons) in the same synthetic realization, and the particle path lengths were compared with the field line path lengths, using both guiding center trajectories and exact test particle orbital  trajectories. 
A systematic difference is found -- the orbital path lengths being larger -- and from this a correction 
factor to account for the additional orbital path length is introduced,
in the form of
an effective pitch angle of the particle 
population as a whole. 

Finally the above approach is implemented to examine path lengths inferred from very recent 
SEP observations by the \isois~instrument suite on Parker Solar Probe \citep{mccomas2016SSR}. A dispersion analysis of six SEP events observed by PSP at heliocentric distance 0.35 AU in late May  and early June 2020 indicates an effective path length of 0.625 AU. To account for this disparity, the FLRW estimation is implemented with the assistance of a global MHD simulation to obtain estimates of local turbulence parameters. This estimate accounts for a little more than half of the added path length inferred from the dispersion analysis. The remainder is accounted for by a plausible effective pitch angle of \(\sim 37\degree\). 

This satisfying result suggests that the easily implemented approach presented here may be useful to provide path length estimates for other SEP events and related observations by PSP and other spacecraft \cite[e.g.,][]{Leske2020ApJS}. Conversely, path lengths inferred from observations of SEP events can potentially provide a measure of magnetic turbulence levels, and possibly complement direct measurements of the magnetic field.

In finalizing this paper it has come to our attention that \citet{LaitinenDalla19apj}
have recently developed a related theoretical approach to computing FLRW influences on SEP path length. In that case the method proceeds by solving a stochastic differential equation for the path length, with a final estimated result that depends on the normalized magnetic field variance. An exact analytical result is available only as an approximation. In the future it will be of interest to compare the results of these two methods, both of which are related to the random walk of magnetic field lines. The related problem of spreading of field lines in the direction transverse to the mean field is considered in Chhiber et al. (2021b, submitted).



\begin{acknowledgements}
We thank Junxiang Hu for useful discussions. This research is partially supported by 
the Parker Solar Probe mission 
and the \isois~project 
 (contract NNN06AA01C) and a subcontract 
 to University of Delaware from
 Princeton University (SUB0000165).
  Additional support is acknowledged from the NASA Living With a Star (LWS) program  (NNX17AB79G) and HSR program (80NSSC18K1210 \& 80NSSC18K1648) and Thailand Science Research and Innovation (RTA6280002).
  The \isois~data and visualization tools are available to the community at \url{https://spacephysics.princeton.edu/missions-instruments/isois}; data are also available via the \href{https://spdf.gsfc.nasa.gov/}{NASA Space Physics Data Facility}. PSP was designed, built, and is now operated by the Johns Hopkins Applied Physics Laboratory as part of NASA’s LWS program (contract NNN06AA01C). Support from the LWS management and technical team has played a critical role in the success of the PSP mission.
\end{acknowledgements}

%
%
\bibliographystyle{aa}
\bibliography{chhibref}
\end{document}